\documentclass [11pt]{article}
\usepackage{setspace,amsfonts,epsfig,amssymb,mathrsfs,amsmath}
\usepackage[margin=1.0in]{geometry}
\usepackage{hyperref}
\usepackage{color}

\newtheorem{axiom}{Axiom}

\pdfoutput=1

\begin{document}

\title{The structure of causal sets\thanks{I wish to thank Fay Dowker for discussions, and Craig Callender and two anonymous referees for comments on an earlier draft. At my own peril, I have not heeded all their advice. This project has been funded in part by the American Council of Learned Societies through a Collaborative Research Fellowship, the University of California through a UC President's Fellowship in the Humanities, and the University of California, San Diego through an Arts and Humanities Initiative Award.}}
\author{Christian W\"uthrich}
\date{For a special issue of the {\em Journal for General Philosophy of Science}, edited by Meinard Kuhlmann and Wolfgang Pietsch.}
\maketitle

\begin{abstract}\noindent
More often than not, recently popular structuralist interpretations of physical theories leave the central concept of a structure insufficiently precisified. The incipient causal sets approach to quantum gravity offers a paradigmatic case of a physical theory predestined to be interpreted in structuralist terms. It is shown how employing structuralism lends itself to a natural interpretation of the physical meaning of causal sets theory. Conversely, the conceptually exceptionally clear case of causal sets is used as a foil to illustrate how a mathematically informed rigorous conceptualization of structure serves to identify structures in physical theories. Furthermore, a number of technical issues infesting structuralist interpretations of physical theories such as difficulties with grounding the identity of the places of highly symmetrical physical structures in their relational profile and what may resolve these difficulties can be vividly illustrated with causal sets. 
\end{abstract}

\begin{center}
\end{center}

\noindent
Re-ignited by John Worrall in 1989 as a response to pessimistic challenges of scientific realism based on historic sequences of false theories,\footnote{Cf.\ e.g.\ Worrall (1989, 117 and 121).} structural realism is currently all the rage in the general philosophy of science. Structural realism---ontically understood---has also risen into prominence as a template to interpret contemporary physical theories over the past decade or so. The former project was, and, to a lesser degree, still {\em is}, plagued by notorious difficulties in explicating the notion of `structure' independently of an {\em ex post facto} identification of structure as that which is preserved through scientific revolutions.\footnote{More recently, Worrall has explicated the structure of a theory in terms of its Ramsey sentence (cf.\ Worrall and Zahar 2001). It is unclear, however, whether this type of structural realism can evade Newman's famous objection (cf.\ Ladyman 2009, \S3.2)} The latter project is typically exhausted by showing that, at least in some respects, the basic constituents of a physical theory are merely structurally identified as they fail to possess some individuality or intrinsic identity. 

Starting out from a rigorously defined set-theoretic characterization of structure in Section \ref{sec:set}, this essay explicates its application to a physical theory, hoping to contribute to an understanding of both these projects. The physical theory to be investigated is causal set theory, an approach to finding a quantum theory of gravity. The main idea behind the causal set approach is to strip spacetime down to its essentials (or less), impose discreteness as an expected quantum signature, and try to obtain a complete theory of quantum spacetime based on these simple assumptions. Even though the approach is still in its infancy and offers less than a complete theory, its conceptual clarity and power elicits admiration and promises to wipe a clean slate when many feel that large parts of established physics will need to be given up in an attempt to obtain a fundamental theory of quantum spacetime. Section \ref{sec:causets} will introduce the main ideas of the theory and discuss its two biggest challenges: to formulate a quantum dynamical law and to understand how general-relativistic spacetimes emerge from the fundamental structure as postulated by causal set theory. 

The reason why causal set theory is of interest to the structuralist is because of its conceptual simplicity and the resulting ease of characterizing the relevant physical structure it postulates and the existence of which the structuralist wants to defend in case a realistic attitude towards it will be warranted. Although others have gestured towards structuralist interpretations of causal set theory, its full explication has hitherto not been offered---it shall be explicated in Section \ref{sec:appl}. In the same section, a recently mounted challenge for spacetime structuralism (W\"uthrich 2009) shall be transposed into the context of causal set theory, and different strategies to address the challenge are evaluated. 

In Section \ref{sec:impl} finally, we shall go for bigger prey and discuss the larger issue of the tenability of structuralist interpretations of scientific theories in the light of the results of Section \ref{sec:appl} and considerations concerning how they may or may not generalize. In particular, this essay draws a sharp distinction between a structuralist interpretation of a particular physical theory on the one hand and structural realism as a wholesale recipe to offer a Third Way between the Scylla of the pessimistic meta-induction and the Charybdis of the no-miracles argument. It will conclude that the foregoing considerations support the former (modulo the difficulty of dealing with causal sets of high degrees of symmetry), but do not at all---or only weakly---support the latter.

\section{The set-theoretical conception of structure}\label{sec:set}

The structuralist analysis of the failure of standard scientific realism in the face of the pessimistic meta-induction concluded that the reason why standard scientific realism did not escape its devastating force was because it insisted on an ontology of `things'.\footnote{Henceforth, I confine myself to {\em ontic} structural realism (cf.\ Ladyman 1998).} In effect, the pessimistic argument was able to unleash its full potency because standard versions of scientific realism read off the surface ontology of objects from the theory only too readily. Consequently, the structuralist concluded, once the realist steers clear of objects and limits her ontological commitment to the deep structure postulated by the theory at stake, then this realist commitment will not be frustrated by a succession of theories with rather different surface ontologies. The problem then, of course, was to independently identify the deep structure a theory postulated. While this issue did not see much progress during the 1990s---and sometimes enjoyed frustratingly scant attention in the literature---, it was usually assumed to be captured by the similarity of the relevant equations, following Worrall (1989).\footnote{More often than not, this assumption was only implicit or, at best, gestured at. A laudable exception is Bain and Norton 2001.} The mathematical structure of equations often does of course grasp the structure of what is postulated by the theory to physically exist, but it should be noted that the mathematics of a theory alone can be very similar for completely dissimilar physical `stuff', while theories trading in more or less the same physical existents may be formulated in terms of rather different mathematics. In fact, one and the same theory may afford distinct mathematical formulations with correspondingly distinct mathematical structures.\footnote{It has been argued that this very fact constitutes an argument in favour of structural realism (Ladyman 1998, 418ff; Bain 2006). While I concur with Jones (1991) that this creates a problem for someone who is predisposed to read off the ontology directly from the formalism, I remain unconvinced that this lends support to structural {\em realism}, for reasons similar to those given by Pooley (2006, \S 4.2).} What ultimately has to matter to the structural realist is the {\em physical} structure towards which we are asked to entertain a realist attitude.

Regardless of what exactly a structure was, structural realist sometimes insisted, and a few continue to insist,\footnote{E.g.\ French (2010).} that their analysis showed that there couldn't fundamentally be any {\em objects}. Instead of fundamental objects, they claimed, ``it's relations all the way down''---in Stachel's (2006, 54) memorable paraphrase of Saunders's (2003, 129) ``it is turtles all the way down''---notably without there fundamentally being any relata to exemplify these relations. More specifically, it is claimed that whatever relata may exemplify the relations ``always turn out to be relational structures themselves on further analysis.''\footnote{Ladyman (1998, \S4). It should be noted that Stachel (2006) does not advocate this position.} Many have complained that such a proposal was simply unintelligible.\footnote{For a representative list, cf.\ Bain (forthcoming, \S2).} This is not the place to rehearse the standard moves along this well-troden path. Suffice it to mention that the standard, set-theoretic conceptualization of structure necessitates the simultaneous postulation of relations as well as their relata.\footnote{It seems to me that a category-theoretic notion of structure in the sense of Bain (forthcoming) is the only promising path to a possible extermination of the relata from what exists fundamentally, even though I remain sceptical that it can ultimately succeed (cf.\ Lam and W\"uthrich 2012).}

Let's explicate this set-theoretic notion of structure, then. The usual, albeit typically tacit, concept of structure assumed in philosophy is that of a {\em relational structure}, i.e.\ a set-theoretic notion found in mathematical logic according to which a structure $\mathcal{S}$ is an ordered pair $\langle O, R\rangle$ which consists of a non-empty set of relations $R$ (`ideology') as well as a non-empty set of relata $O$ (`ontology'), the {\em domain} of $\mathcal{S}$. More generally, a structure consists of a set of elements together with a collection of finitary functions and relations defined on the set. More precisely, a {\em structure} $\mathcal{A}$ is an ordered triple $\langle A, R_n, F_n \rangle$ consisting of a domain (or universe) $A$, a (usually countable) set $R_n$ of $n$-ary relations, and a (usually countable) set $F_n$ of $n$-ary functions.\footnote{Even more precisely, given that the distinction in mathematical logic between relations and functions on the one hand and symbols representing them on the other, a structure can be defined as the ordered 5-tuple $\langle A, \bar{R}_n, \bar{F}_n, \mbox{ar}, I\rangle$ consisting of a domain $A$, a set of $n$-ary relations symbols $\bar{R}_n$, a set of $n$-ary function symbols $\bar{F}_n$, a function $\mbox{ar}: \bar{R}_n\cup \bar{F}_n \rightarrow \mathbb{N}_0$ which assigns a natural number (including $0$) called {\em arity} to every symbol in $R_n\cup F_n$, and an interpretation function $I$ assigning functions (and constants) and relations to the symbols in $\bar{F}_n$ and $\bar{R}_n$. In what follows, I shall speak loose and fast in that no distinction will be made between a symbol $\sigma$ and its interpretation $I(\sigma)$.} The {\em domain} $A$ of $\mathcal{A}$, sometimes also denoted $\mbox{dom}(\mathcal{A})$, is just an arbitrary, usually non-empty, set. A function $f$ is {\em $n$-ary} just in case there exist sets $X_1,..., X_n, Y$ such that $f: X_1\times \cdots \times X_n \rightarrow Y$, where $X_1 \times \cdots \times X_n$ is the Cartesian product of $X_1,..., X_n$. The Cartesian product $X_1 \times \cdots \times X_n$ of $X_1,..., X_n$ is defined as the set of all ordered $n$-tuples $\langle x_1,..., x_n\rangle$ such that for all $i=1,...,n$, $x_i \in X_i$. An {\em $n$-ary relation} defined on sets $X_1,..., X_n$ is a set of ordered $n$-tuples $\langle x_1,..., x_n\rangle$, where $x_i\in X_i$ for all $i=1,...,n$. Thus, an $n$-ary relation on sets $X_1,...,X_n$ is just a subset of the Cartesian product $X_1\times\cdots\times X_n$ of these sets.\footnote{\label{fn:redfuncs} For those with a penchant for ever higher levels of generality, ($n$-ary) functions can also be defined as ($(n+1)$-ary) relations.} Without loss of generality (see footnote \ref{fn:redfuncs}), we confine ourselves to {\em relational structures}, i.e.\ to structures with $F_n = \emptyset$.

This standard notion has recently received critical scrutiny from various authors. Leitgeb and Ladyman (2008) explore a graph-theoretic notion of structure,\footnote{Note, however, that a graph-theoretic rendering of what it is to be a structure may reasonably be considered a form of {\em representing} a set-theoretically defined structure, rather than offering an alternative, and inequivalent, characterization. For more discussion of structures and their representations, cf.\ Lam and W\"uthrich (2012).} and Roberts (2011) proposes a group-theoretic one. Landry (2007) argues for conceptual pluralism in that a restriction to the standard notion (or any {\em one} notion, for that matter) cannot do justice to the plethora of applications. Muller (2010) makes the case that a thoroughgoing structuralism should not rely on a conceptualization of structure that rests on a definition in terms of more basic concepts. Instead, he argues, structure should be given as a primitive, encoded in an axiomatization. Since the set-theoretic notion, however, is both standard and adequate given my purposes, the focus shall be on it in what follows.

For our work below, it will become important to be able to compare structures. In particular---and this also matters to the structural realist who wants to escape the pessimistic meta-induction by identifying structures that are preserved through successions of theories---, we need a notion of structural identity, i.e.\ we need some criteria as to when two structures are `identical'. Since we are interested in {\em physical} structures, and these structures can only be {\em in re},\footnote{As opposed to {\em ante rem}, to use Shapiro's (1997) distinction. {\em In re} structuralism maintains that the physical system which exemplifies a certain structure is ontologically prior to the abstract mathematical structure, i.e.\ the isomorphism class of `identical' structures. In contrast, {\em ante rem} structuralism maintains that abstract structures exist and are ontologically prior to any physical exemplification of it (or at least ontologically independent of them).} we need a conceptual apparatus to articulate how two physical `things' existing in the same possible world, or how the physical existents in two different possible worlds (or sets thereof), can have the `same' structure. For instance, the inter-mundial version of structural identity is required to determine whether the physically possible worlds according to one theory are structurally identical to those worlds possible according to its successor, or at least partially so. In other words, it appears to be needed to precisify the structural continuity that a structural realist asserts of successive scientific theories. 

Roughly, a `homomorphism' is a `structure-preserving' map from one structure to another. If a bijective homomorphism has an inverse that is also a homomorphism, then we say that it is an `isomorphism'. Isomorphisms are usually used to capture the structural identity we are seeking: two structures $\mathcal{A}$ and $\mathcal{B}$ are {\em structurally identical} just in case they are isomorphic, denoted $\mathcal{A} \simeq \mathcal{B}$, i.e.\ iff there is an isomorphism from $\mathcal{A}$ to $\mathcal{B}$. Note that this sense of identity is weaker than that of (strict) identity, since in general $\mbox{dom}(\mathcal{A}) \neq \mbox{dom}(\mathcal{B})$. An isomorphism from a set onto itself is an {\em automorphism}. In this case, the domains are identical. 

To state all of this somewhat more rigorously, suppose we are given two structures $\mathcal{A}$ and $\mathcal{B}$ with the same set of relations defined over their (generally distinct) domains.\footnote{An attentive reader may wonder how purely extensionally defined relations may be the `same' if defined over two different sets. This difficulty is circumvented in the more precise language of mathematical logic where symbols denoting relations are distinguished from the relations they denote. Strictly speaking, therefore, the {\em signatures} of the two structures (i.e.\ their sets of relation symbols) must be the same, but not, of course, the interpretation of these symbols.} A {\em homomorphism} from $\mathcal{A}$ to $\mathcal{B}$ is a map $\phi:A\rightarrow B$ which preserves the functions and relations, as follows: (i) for any $n$-ary function $f$ in $F_n$ and any elements $a_1,...,a_n\in A$, $\phi(f(a_1,...,a_n)) = f(\phi(a_1),..., \phi(a_n))$; and (ii) for any $n$-ary relation $R$ in $R_n$ and any elements $a_1,...,a_n\in A$, $(a_1,...,a_n) \in R_n \Rightarrow (\phi(a_1),...,\phi(a_n)) \in R_n$. A bijective map $\phi: A\rightarrow B$ is called an {\em isomorphism} just in case both $\phi$ and its inverse $\phi^{-1}$ are homomorphisms. Two structures $\mathcal{A}$ and $\mathcal{B}$ are called {\em isomorphic}, denoted by $\mathcal{A}\simeq \mathcal{B}$, or structurally identical, iff there exists an isomorphism $\iota: A \rightarrow B$. If $\mathcal{A}\simeq\mathcal{B}$ and $A=B$, then we say that $\mathcal{A}$ and $\mathcal{B}$ are {\em automorphic} and the relevant isomorphism is called an {\em automorphism}.

This concludes our brief statement of the set-theoretic notion of structure. We are now ready to see it applied in a concrete case.

\section{Introducing causal sets}\label{sec:causets}

The relevance of quantum effects and the presence of strong gravitational fields in the early universe as well as in black holes necessitates a quantum theory of gravity, i.e.\ a theory that combines (possibly classical) gravity with quantum theories of matter in order to describe their interaction. To many---though not to all---, this means to quantize gravity, i.e.\ to formulate a theory that either results from some regimented quantization procedure applied to the classical gravitational field or directly relies on postulates which codify the quantum structure of the gravitational field or of spacetime. The causal set theory attempts the latter: it encodes in its two kinematic axioms the basic idea of the quantum analogue of spacetime as something essentially discrete that is structured by causal relations and hopes to show that at appropriately large scales, this discrete quantum structure approximates the smooth metric manifolds that represent spacetime in general relativity. Thus, the founding principle of causal set theory is the idea that the fundamentally discrete structure consists of a partially ordered set of elementary events, and that this ordering is essentially causal.\footnote{The original paper that got everything started is Bombelli et al.\ (1987), even though there are clear historic precursors in Myrheim (1978) and 't Hooft (1979). For serviceable reviews of the causal sets approach, cf.\ Dowker (2005), Henson (2009), and Reid (2001). As far as I am aware, the philosophical reception is exhausted by brief discussions in Butterfield (2007, \S 5.2), Earman (2008, \S 7), Smeenk and W\"uthrich (2011, \S 8), and Stachel (2006, \S 3.7).}

Even though the discreteness and the fundamental role of causality are stipulated {\em ab initio} in causal set theory, it is of course not the case that this stipulation is unmotivated. Important results in the 1970s made it clear that in general relativity, the causal structure of a spacetime determines its geometry (but not its `size'). Theorems by Hawking et al.\ (1976) and Malament (1977) established that given the causal structure and volume information, one finds the dimension, the topology, the differential structure, and the metric of the manifold. These results kindle the hope that minimal assumptions concerning causality at the fundamental level may suffice to recover almost all the requisite structure we care about in general-relativistic spacetimes and thus motivate the causal sets approach in assuming that the fundamental structure is a `causal set'. 

The discreteness is justified, advocates of the approach claim, by both physical as well as technical reasons. Among the latter, it is argued (in Henson 2009, 394) that the lack of short-distance cut-offs in the relevant degrees of freedom leads to a host of infinities in general relativity and quantum field theory and that at least those infinities not resolved by renormalization are best cured by assuming a smallest spatial distance. But these difficulties---at least in general relativity---simply result from insisting on forcing general relativity on the Procrustean bed of perturbation theory, an insistence dropped in so-called background-independent approaches such as loop quantum gravity and causal dynamical triangulations. Similarly, technical problems in defining path-integrals on a continuous history space can be avoided by assuming fundamental discreteness. Quite generally, Henson (ibid.) argues, discreteness affords a conceptual utility in that it often significantly reduces the technical complexities physicists face. But without a metaphysical commitment to the simplicity of nature, there is no reason to think that the actual world is fundamentally discrete just because assuming so mitigates our mathematical struggles. 

But there are also physical motivations for believing that the fundamental structure may be discrete. First, what Henson considers ``perhaps the most persuasive argument'' (ibid.) is that without a short-distance cut-off, the finiteness of the semi-classical black hole entropy cannot be obtained. For this argument to be persuasive, however, the importance of keeping the black hole entropy finite must be accepted. Suffice it to point out that the Bekenstein formula for the black hole entropy is derived in {\em semi-classical} quantum gravity, i.e.\ by mixing and matching physical principles of which we cannot be certain will be licensed by a full quantum theory of gravity. To be sure, the confluence of independent lines of reasoning to Bekenstein's functional expression for the entropy of black holes is suggestive. But this argument from black hole entropy to discreteness could be turned around; i.e.\ it could be argued that given that the full quantum theory of gravity trades in a {\em continuous} fundamental structure, Bekenstein's formula cannot be right. A second physical reason is offered by Reid (2001, 6): the local conservation of energy suggests that photons with infinite energies do not exist and an effective way of ascertaining this is by assuming that time---and therefore, by relativity, space---is discrete. No doubt this is a possibility; but in itself, it falls short of offering a conclusive argument.

The most puzzling justification for discreteness comes also from Henson (ibid.). Since various approaches to quantum gravity assume, suggest, or entail that the fundamental structure is discrete, he argues, it is only reasonable for causal set theory to postulate discreteness. But of course, these competing theories will not be true if causal set theory is; so why should this lend any credence to causal set theory? Perhaps the idea is to show that given the agreement of rather diverse programmes in that the fundamental physical structure is discrete, we should be confident in what they concur on. But since the discreteness is a stipulated a priori in causal set theory, it is certainly not the case that the causal set programme adds any support to the hypothesis of discreteness. But this brings us back to the problem how this assumption is justified via an inter-theoretic agreement unless at least one of the competing theories is true. Either way, then, if we take the causal set programme's ambition to be the providing of a full quantum theory of gravity, then any inter-theoretic concurrence about the discreteness of the fundamental structure cannot offer any support for the programme's stipulation. But perhaps such ambition is not harboured by the programme and it merely strives to give a conceptually simple and clear line of attack to gain some understanding of potential features of a full theory. If therefore no claim as to the truth of its basic premises is implied, then this convergence can act as evidence for its physical relevance and thereby justify interest in this approach. 

Hence, none of these arguments offers direct evidence for discreteness. But this may not deter us as ultimately, of course, the postulates of a theory are justified by the theory's success. If the theory aspires to be true and fundamental, only unconditional and complete empirical confirmation will be able to elevate it to that status. If the theory is less ambitious and just aims to provide a toy theory which usefully models some features that a more complete and fundamental theory is expected to have, then it just needs to establish that these features reappear indeed in the fuller theory and lend credence, either via theoretical or empirical support, to its claim to model them with relevant similarity. 

\subsection{The kinematics of causal set theory}\label{ssec:kin}
The fundamental structure is a {\em causal set} $\mathcal{C}$, i.e.\ an ordered pair $\langle C, \preceq\rangle$ consisting of a set $C$ of elementary events and a relation, denoted by the infix $\preceq$, defined on $C$ satisfying the following conditions:

\begin{axiom}[Partial ordering]\label{ax:partial}
$\preceq$ induces a {\em partial order} on $C$, i.e.\ it is reflexive ($\forall a \in C, a \preceq a$), antisymmetric ($\forall a, b \in C,$ if $a \preceq b$ and $b \preceq a$, then $a=b$), and transitive ($\forall a, b, c \in C$, if $a \preceq b$ and $b \preceq c$, then $a \preceq c$).
\end{axiom}

\begin{axiom}[Local finitarity]\label{ax:discrete}
$\forall a, c \in C, {\bf card}(\{b \in C | a \preceq b \preceq c\}) < \infty$.
\end{axiom}
These two simple axioms determine the kinematics of causal set theory and have been motivated above. Axiom \ref{ax:discrete} encodes the discreteness of causal sets. Interpreting the relation $\preceq$ causally, Axiom \ref{ax:partial} precludes any causal loops. In other words, time travel is ruled out by stipulation. This may or may not be physically costly, but it should be noted that however causal set theory will relate back to general relativity, it will not be able to reproduce the full theory as general relativity permits time travel in the sense of causal loops (cf.\ Smeenk and W\"uthrich 2011). 

\begin{figure}
\centering
\epsfig{figure=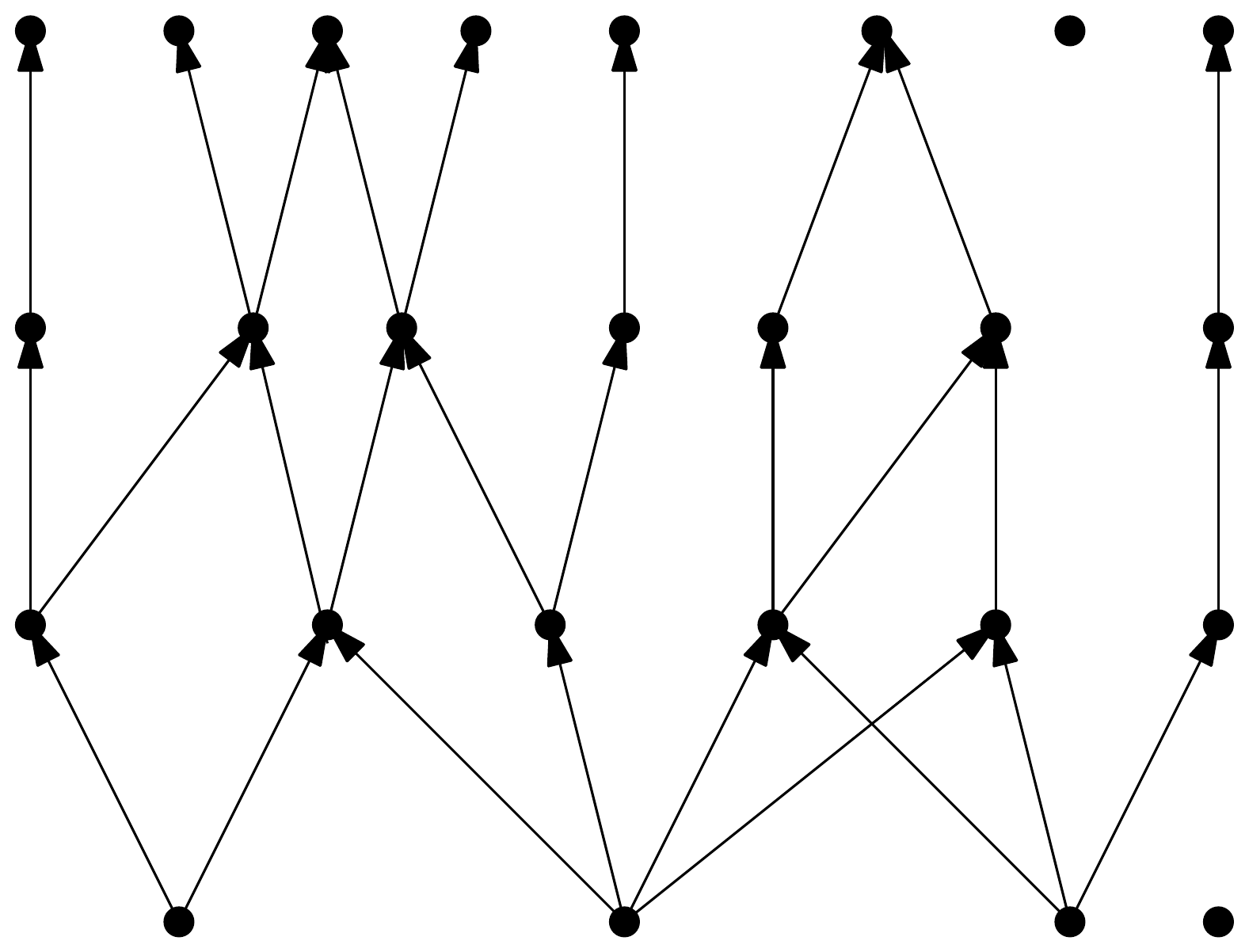,width=0.5\linewidth}
\caption{\label{fig:hasse} (Almost) a Hasse diagram of a causal set}
\end{figure}

Causal sets can be illustrated by so-called {\em Hasse diagrams}, graph-theoretic representations of finite partially ordered sets. In these diagrams, the vertices represent elements of $C$ and edges connecting the vertices represent their standing in relation $\preceq$. Since $\preceq$ is reflexive and transitive, only a causal set's `transitive reduction' is drawn, i.e.\ we do not connect vertices directly which are already connected via intermediate vertices.\footnote{\label{fn:trans}The {\em transitive closure} of a binary relation $R$ on a set $X$ is the intersection of all transitive relations on $X$ that contain $R$. The {\em transitive reduction} of a binary relation $R$ on a set $X$ is the smallest relation on $X$ which has the same transitive closure as $R$.} Furthermore, since it is understood that $\preceq$ is reflexive, we also don't clog the diagrams by drawing edges from each vertex back to itself. An example of a simple causal set is represented in Figure \ref{fig:hasse}. This is strictly speaking not a Hasse diagram as its edges are adorned with arrowheads, which have been introduced in the present context to emphasize the fact that the relation they represent is antisymmetric. 

\subsection{Steps towards a dynamics of the theory}\label{ssec:dyn}
\begin{figure}
\centering
\epsfig{figure=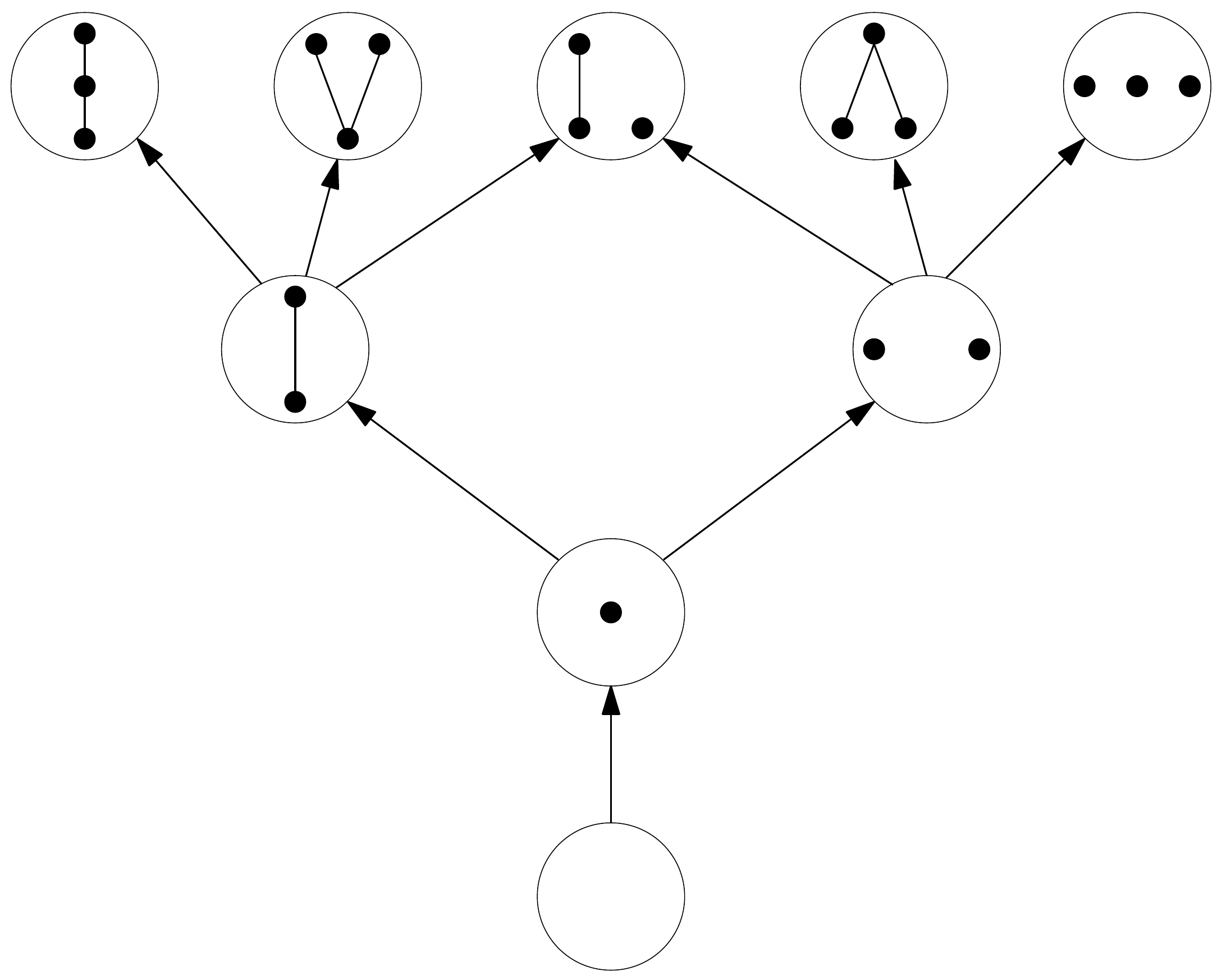,width=0.5\linewidth}
\caption{\label{fig:dyn} The partially ordered set of finite causal sets (with the arrowheads within each causal set now omitted)}
\end{figure}

Even though the basic kinematic rules in \S\ref{ssec:kin} give essentially the complete structure of causal sets, they don't give the full theory. Causal sets are dynamical entities in that they grow by the `birthing' of additional elements to the causal future of the existing causal set. Among different proposed dynamical rules, the {\em classical sequential growth dynamics} proposed by Rideout and Sorkin (1999) has been most influential.\footnote{See also Varadarajan and Rideout 2006.} In this dynamics, causal sets `grow' by a discretized, stochastic Markov process.\footnote{A {\em Markov process} is a dynamical process whose transition amplitudes are such that the corresponding probabilities satisfy a Markov condition, which states, roughly, that the probabilities for the transitions only depend upon the `initial' and the `final' state, but not on what transpired before the `initial' state.} This growth process starts out from the empty set and adds, one by one, new elements to the future of the existing ones, following a directed path in Figure \ref{fig:dyn}, which shows all possible routes an evolution can take. More precisely, if an auxiliary, i.e.\ non-physical, external time is introduced to label the events in order of their birth, then the events' labels $l\in \mathbb{N}_0$ must obey the condition that if $x\preceq y$, then $l(x) \leq l(y)$. This demand is called {\em internal temporality}. It is important to note that the converse implication does not hold, as new events may be spacelike related to all existing ones. 

This labelling will introduce representational surplus structure in that e.g.\ the causal set in the centre of the top line in Figure \ref{fig:dyn} can be reached via two different paths and hence can occur with two inequivalent labellings.\footnote{More precisely, {\em three} different labellings, since the third event could can related to either of the spacelike related events in the causal set on the right of the ${\bf card}(C) = 2$ generation.} Causal set theorists codify their demand that the labels carry no physical meaning in the requirement for {\em discrete general covariance}, which asserts that the transition probability between two causal sets related dynamically as along a directed path as in Figure \ref{fig:dyn} is path-independent insofar as the product of the transition probabilities along each path must be the same. Discrete general covariance ensures that the labels used in the growth sequence are `pure gauge'.\footnote{Cf.\ Rideout and Sorkin (1999), Brightwell et al.\ (2002).} This means that the true dynamical content of the theory is contained in, and exhausted by, the transition amplitudes and not---emphatically---in the path along which a causal set is thought to grow.

The structure $\mathcal{P}$ consisting of the set of finite causal sets representing all different dynamical stages a growing causal set can assume is again partially ordered. The ordering relation is not a causal relation as it was at the kinematic level, but instead a {\em parent-child relation} which takes finite causal sets as it relata. Two finite causal sets are related by the parent-child relation just in case one can be obtained form the other by accreting an event to the latter in accordance with internal temporality. The structure $\mathcal{P}$ can again be represented by a Hasse diagram, as shown in Figure \ref{fig:dyn}. It should be clear that $\mathcal{P}$ has a first element---the empty set---, but no last element. 

Furthermore, a condition of relativistic causality is typically imposed. Without going into any technical detail here, the idea behind such a `Bell causality' assumption is that what transpires at spacelike separation from the node to which the newborn event is cemented ought not to influence the corresponding transition amplitude. It can be shown that, together with discrete general covariance, an intuitive notion of `Bell causality' imposes strong constraints on the dynamics.\footnote{For the technical results, see Rideout and Sorkin (1999) and Varadarajan and Rideout (2006); for their foundational appraisal, see Butterfield (2007).}

All of this remains entirely classical, even though stochastic. No pretension is made to model effects arising from quantum interference. No superpositions of causal sets are considered. Surely, this gives us yet another reason for believing that causal set theory is not a serious contender for a full quantum theory of gravity---at least not so long as no genuinely quantum dynamics is offered. Researchers in causal set theory, of course, admit as much and often consider classical sequential growth dynamics as a `warmup exercise'. 

The classical sequential growth dynamics is just one among many possibilities to implement a dynamical evolution in the theory. In an attempt to curb the number of possibilities, it, like others, imposes constraints such as discrete general covariance and a Bell-type causal condition. These constraining principles certainly appear eminently reasonable, but they lead to a serious problem: it seems as if they prescribe dynamics that tend {\em not} to result in causal sets which give rise to manifold-like spacetimes (Georgiou 2005). Let us address the issue of the emergence of classical spacetimes from causal sets then.

\subsection{The emergence of classical spacetimes from causal sets}\label{ssec:emerge}
For any quantum theory of gravity to succeed, it must be able to recover the classical spacetimes of general relativity in the appropriate low-energy limit. Only if such a theory can show how classical spacetimes emerge in the relevant sense---and thereby explain why general relativity is as successful as it is---does it become a contender to supplant general relativity. Thus, studying the relationship between discrete causal sets and continuum spacetimes constitutes an integral part of the approach.\footnote{The term `emergence' is not intended in the philosophers' sense in which it designates a not even weak reducibility; rather, it is used in the physicists' sense of an umbrella term for reducibility broadly understood.}

What we would like to know, more specifically, is whether there exists an embedding of a given causal set $\langle C, \preceq\rangle$ into some relativistic spacetime $\langle \mathcal{M}, g_{ab}\rangle$, i.e.\ whether there is a map $\phi: C \rightarrow \mathcal{M}$.\footnote{For a recent review of the problem of the emergence of space in particular from causal sets, see Rideout and Wallden 2009.} This embedding should be {\em faithful} in that it satisfies the following three conditions:
\begin{enumerate}\renewcommand{\labelenumi}{(\roman{enumi})}
\item The causal relations are preserved, as follows: $\forall a,b \in C, a\preceq b$ iff $\phi(a) \in J^-(\phi(b))$, where $J^-(p)$ is the {\em causal past} of $p$, i.e.\ the set of points $q\in\mathcal{M}$ such that there exists a future-directed causal curve from $q$ to $p$ (or $q=p$).
\item On average, $\phi$ maps one element of $C$ onto each Planck-sized volume of $\langle\mathcal{M}, g_{ab}\rangle$.
\item $\langle\mathcal{M}, g_{ab}\rangle$ has no length scales smaller than the `discreteness scale' of the causal set; for instance, it is approximately flat below the Planck scale (the typical discreteness scale).
\end{enumerate}
If the third condition is violated, then the relativistic spacetime has features that the causal set cannot hope to retrieve---it simply lacks the resources to incorporate below Planck-scale phenomena. If there is a faithful embedding of a causal set into a relativistic spacetime, it is common to say that the causal set at stake is {\em manifold-like}. 

The hope now is that if a faithful embedding of the causal set into a relativistic spacetime exists, it is essentially unique insofar as if $\phi': C\rightarrow \mathcal{M}'$ is a faithful embedding into another relativistic spacetime $\langle \mathcal{M}', g_{ab}'\rangle$, then the two spacetimes are approximately isometric, at least at scales above the discreteness scale. This hope, captured in what is often referred to as the `causal set hauptvermutung', has so far not been substantiated by proof. 

Moreover, as it turns out it is far from trivial to determine whether a given causal set indeed gives rise to a relativistic spacetime. Because of this, let us consider the reverse direction from relativistic spacetimes to causal sets in the hope to glean some lessons that may apply to the original direction. In considering how to find a manifold-like causal set starting from a continuum spacetime, we first notice that the general recipe to do this is by `sprinkling' points onto the manifold approximately in proportion to the volume of the spacetime regions. Next, the causal structure of the spacetime is used to determine ordering relations among the sprinkled points, i.e.\ to establish the precedence relation $\preceq$ among the sprinkled points. Using this recipe, we are guaranteed to obtain a causal set that can be faithfully embedded into the spacetime from which we started out. 

In order for the embedded causal set to be invariant under Lorentz transformations, not just any old way of sprinkling will do. In particular, regular sprinkling patterns such as Planck-length-spaced lattices will not be Lorentz invariant as the boosted lattice may have strongly varying point densities across the manifold (Dowker et al.\ 2004). The way to protect the embedding against a loss of Lorentz invariance is by sprinkling the points randomly. Causal set theory uses a Poisson process where the probability of sprinkling $N$ points into a spacetime volume is given by a Poisson distribution. Since Poisson processes in $\mathbb{R}^n$ are invariant under any volume-preserving map and the Minkowski volume element equals the Euclidean volume element in $\mathbb{R}^n$, the Poisson sprinkling exhibits exact Lorentz invariance for Minkowski spacetime (Dowker 2005, 451). Since the volume element of general-relativistic spacetimes will not equal that of Euclidean space, the Lorentz invariance of the sprinkling will in general only hold locally.

The question of the emergence of spacetimes from causal sets is riddled with what Smolin (2006, 211) has dubbed the {\em inverse problem for causal sets}, viz.\ the fact that almost no causal sets can be faithfully embedded into a relativistic spacetime. This may in itself not constitute a severe problem, {\em if} we had some physically well-motivated principle that permitted the selection of the subset of the manifold-like causal sets among all kinematically or dynamically possible ones. As I mentioned in \S \ref{ssec:dyn}, classical sequential growth dynamics fails to offer a dynamical mechanism that would drive the causal sets to grow such as to render them manifold-like. Generally, causal set theory does not have the resources to select in a principled way, or a dynamical principle to generate, those causal sets that approximate low-dimensional relativistic spacetimes.\footnote{\label{fn:dyn}The role of time and of the dynamical principles considered in causal set theory is a subtle business and will have to be analyzed on another occasion.}

\section{A structuralist interpretation of causal set theory}\label{sec:appl}

It is straightforward to interpret the causal sets postulated by causal set theory in structuralist terms, i.e.\ as structures.\footnote{I am not the first to notice: cf.\ Brightwell et al.\ (2002, 8), Smolin (2006, \S 7.2.1), and Stachel (2006, \S3.7). It should be noted however, that the first two don't make explicit use of the label `structuralism', and the last one only offers a rather concise assertion to this effect.} The domain $O$ of the structure $\mathcal{C}$ is $C$, the set of elementary events, and the set $R$ of relations defined on the domain is exhausted by $\preceq$. For the structuralist, it matters that the elements of $C$ do not have intrinsic individuality---they are completely featureless events which assume their identity merely by virtue of the position they occupy in the structural complex $\mathcal{C}$. The relation $\preceq$ is the only concrete physical relation. The structure of causal sets is completely exhausted by the kinematic axioms Axiom \ref{ax:partial} and Axiom \ref{ax:discrete}, perhaps amended by the dynamical rules specified in \S\ref{ssec:dyn}. It is thus evident that causal sets offer what is arguably the most straightforwardly structuralist example of a physical entity postulated by any physical theory. 

Of course, structuralist interpretations of the physics of other theories have been proposed.\footnote{E.g.\ general relativity, non-relativistic quantum mechanics, and quantum field theory; cf.\ Ladyman and Ross (2007, Ch.~3), Esfeld and Lam (2008, 2011).} In particular, it has been claimed that structuralism (or structural realism) about spacetime naturally dissolves the indeterminist trap into which a (naive) substantivalist stepped in the face of the hole argument in general relativity, while still maintaining a form of realism about spacetime, thereby avoiding a number of difficulties for the relationalist. According to the structuralist, spacetime ought to be interpreted as a physical structure, i.e.\ as a structural complex constituted by physical objects---the relata---and by the concrete physical relations in which they stand. Usually, this is rendered as taking the points of the manifold $\mathcal{M}$ as the set of physical objects and the metrical, i.e.\ spatio-temporal, as well as the dynamical properties they exemplify as the concrete physical relations. As articulated, e.g., in Esfeld and Lam (2008, \S2), spacetimes are thus identified with an equivalence class of diffeomorphically related models of general relativity. The hole argument trades on the intrinsic individuation of manifold points prior to, and independent of, the spatio-temporal relations in which they stand. Such individuation is what forces the conclusion that diffeomorphically-related models of general relativity must represent genuinely distinct physical situation; and it is exactly this individuation which is denied by (all versions of) ontic structural realism about spacetime. 

It has been shown (W\"uthrich 2009) that the important family of relativistic cosmological models called Friedmann-Lema\^{\i}tre-Robertson-Walker (FLRW) spacetimes challenge this structuralist interpretation of spacetime. An important feature of these spacetimes is that the four-dimensional spacetime permits a privileged foliation into three-dimensional spaces parametrized by a one-dimensional so-called {\em cosmological time}. Another important feature of them is that they encode the demand issued by the `Cosmological Principle' that, at any time, there be no privileged spatial location in the universe. In other words, there cannot be any physical property had by any point in the universe that is not also had by all other points at the same cosmological time. If a point cannot differ in its properties, including its relational properties, from any other point in space, then the `Principle of the Identity of Indiscernibles' (PII) appears to demand that they be identified. PII holds that for any two objects, if they share all---possibly relational---properties, then they are identical. If sound, this argument establishes that the spacetime structuralist is committed to the absurd claim that FLRW spacetimes consist of only one point for any given value of cosmological time---a pointlike universe!\footnote{\label{fn:primid}There are several ways to evade the strictures of this argument, as explained in W\"uthrich 2009. One could, of course, assert that identity (or at least numerical plurality) is primitive, as do e.g.\ Leitgeb and Ladyman (2008), thus rendering the application of PII otiose. I wish to thank an anonymous referee for keeping me honest on this point. I shall assume, in what follows, that identity is contextual rather than primitive.}

It has been suggested, most vigorously in Muller 2011, that the introduction of irreflexive relations resolves the difficulty insofar as these relations render the relata {\em weakly discernible}, i.e.\ the relata are numerically distinct by virtue of them exemplifying an irreflexive relation.\footnote{A binary relation $R$ on a set $X$ is {\em irreflexive} just in case $\forall x\in X, \neg Rxx$.} The symmetry problem disappears since for an irreflexive relation to be exemplified at all, there must be two numerically distinct objects. There is the general worry with this resolution, of course, that the assumption of there being an irreflexive relation exemplified in the physical structure at stake means, {\em eo ipso}, that there are two numerically distinct objects exemplifying the relation.\footnote{To claim that weak discernibility may be used to introduce an identity relation to a language without identity amounts to an insistence that the pertinent relation is {\em primitively} irreflexive. Thus, just as in footnote \ref{fn:primid}, such a claim is tantamount to accepting non-structural facts, a possibility I acknowledge without pursuing it here.} If the point of this resolution was that numerical plurality was to be {\em derived}, rather than stipulated, then it seems to fail.

In causal set theory, the only physically admissible fundamental relation is reflexive, thus not directly amenable to the move proposed by Muller. But it is straightforward to concoct an irreflexive relation from a reflexive one. First, define the {\em reflexive reduction} $R^{\neq}$ of a binary relation $R$ over a set $X$ as $R \setminus \{ \langle x, x\rangle | x \in X\}$, i.e.\ the reflexive reduction $R^{\neq}$ of a possibly reflexive relation $R$ is exemplified by exactly the same pairs of objects as $R$ except for those pairs when the objects in both slots of the binary relation are identical. In fact, some authors contributing to the physics literature on causal sets characterize the fundamental relation exactly as the reflexive reduction of the one defined in Axiom \ref{ax:partial}.\footnote{\label{fn:irrefl}Cf.\ e.g.\ Dowker et al.\ (2004).} 

Second, convince yourself that not much physically hangs on whether the fundamental relation is reflexive or irreflexive. To this end, define a {\em strict partial order} as a binary relation that is irreflexive and transitive, and hence asymmetric.\footnote{A binary relation $R$ on a set $X$ is {\em asymmetric} just in case $\forall x, y\in X$, if $Rxy$, then $\neg (Ryx)$.} If we denote {\em strictly ordered causal sets} ordered pairs $\langle C, \prec\rangle$ of sets $C$ of elementary events endowed with a strict partial ordering relation, denoted by the infix $\prec$, and satisfying Axiom \ref{ax:discrete}, then there is a bijective correspondence between causal sets and strictly ordered causal sets. To see this, suppose the set $C$ is the same for both. Then given the reflexive relation $\preceq$ inducing a non-strict partial order on $C$, we can always introduce an irreflexive relation $\prec$ given by the  reflexive reduction $\preceq^{\neq}$ such that $\forall x, y \in C, x\prec y$ just in case $x\preceq y$ and $x\neq y$. Conversely, given the irreflexive relation $\prec$ inducing a strict partial order on $C$, we can always introduce a reflexive relation $\preceq$ given by the reflexive closure $\prec^=$ such that $\forall x, y\in C, x\preceq y$ just in case $x\prec y$ or $x=y$. Generally, the {\em reflexive closure} $R^=$ of a binary relation $R$ over a set $X$ is defined as $\{\langle x, x\rangle | x\in X\} \cup R$, i.e.\ the reflexive closure $R^=$ of a possibly irreflexive relation $R$ is exemplified by those pairs of objects which either exemplify $R$ or take identical objects in both slots.

It is clear that a causal set $\mathcal{C}=\langle C, \preceq\rangle$ cannot be isomorphic to its strictly partially ordered counterpart $\mathcal{C}^s=\langle C, \prec\rangle$ since no map between them will preserve the one and only relation of the structure either way. Thus, strictly speaking, causal set theory and `strict causal set theory' are inequivalent as they describe non-isomorphic structures. But to conclude from this that there is any physically interesting difference between either these two theories or the structures they describe would nevertheless be misguided as the theories are still empirically equivalent.\footnote{Exercise for the reader: try to exploit the difference between causal set theory and its strict analogue to define an empirically tangible difference of the emergent spacetimes that they engender.} It appears to make good empiricist sense to be indifferent regarding a possible preference between causal set theory and its strict cousin. Thus, causal set theorists might as well have formulated the entire theory in terms of a basic relation that is irreflexive.\footnote{And in fact, some do: cf.\ footnote \ref{fn:irrefl}.}

\begin{figure}[h]
\hfill
\begin{minipage}[t]{0.45\textwidth}
\begin{center}  
\epsfig{figure=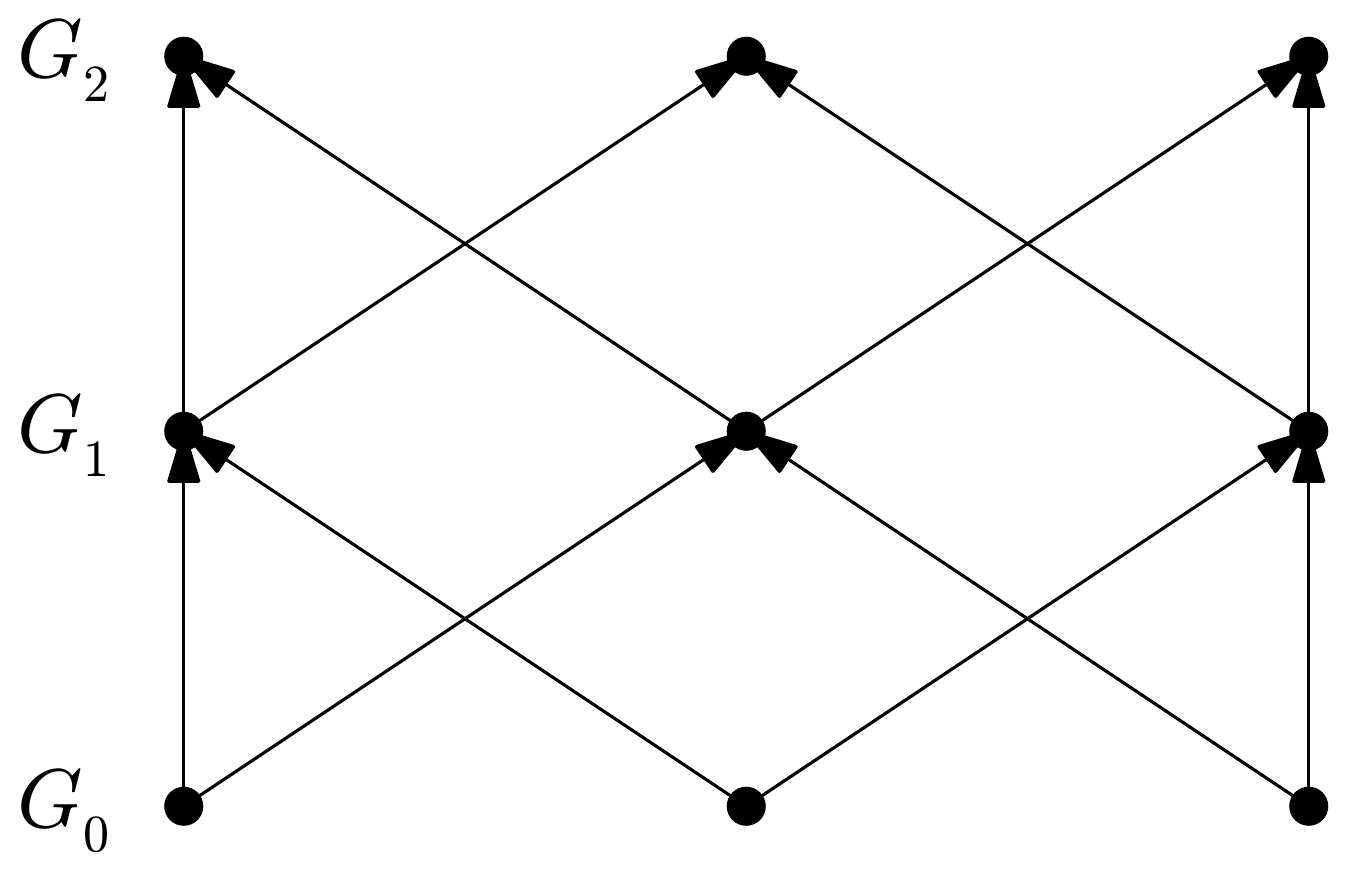,width=0.95\linewidth}
\caption{A symmetric causal set}
\label{fig:symcauset}
\end{center}
\end{minipage}
\hfill
\begin{minipage}[t]{0.45\textwidth}
\begin{center}  
\epsfig{figure=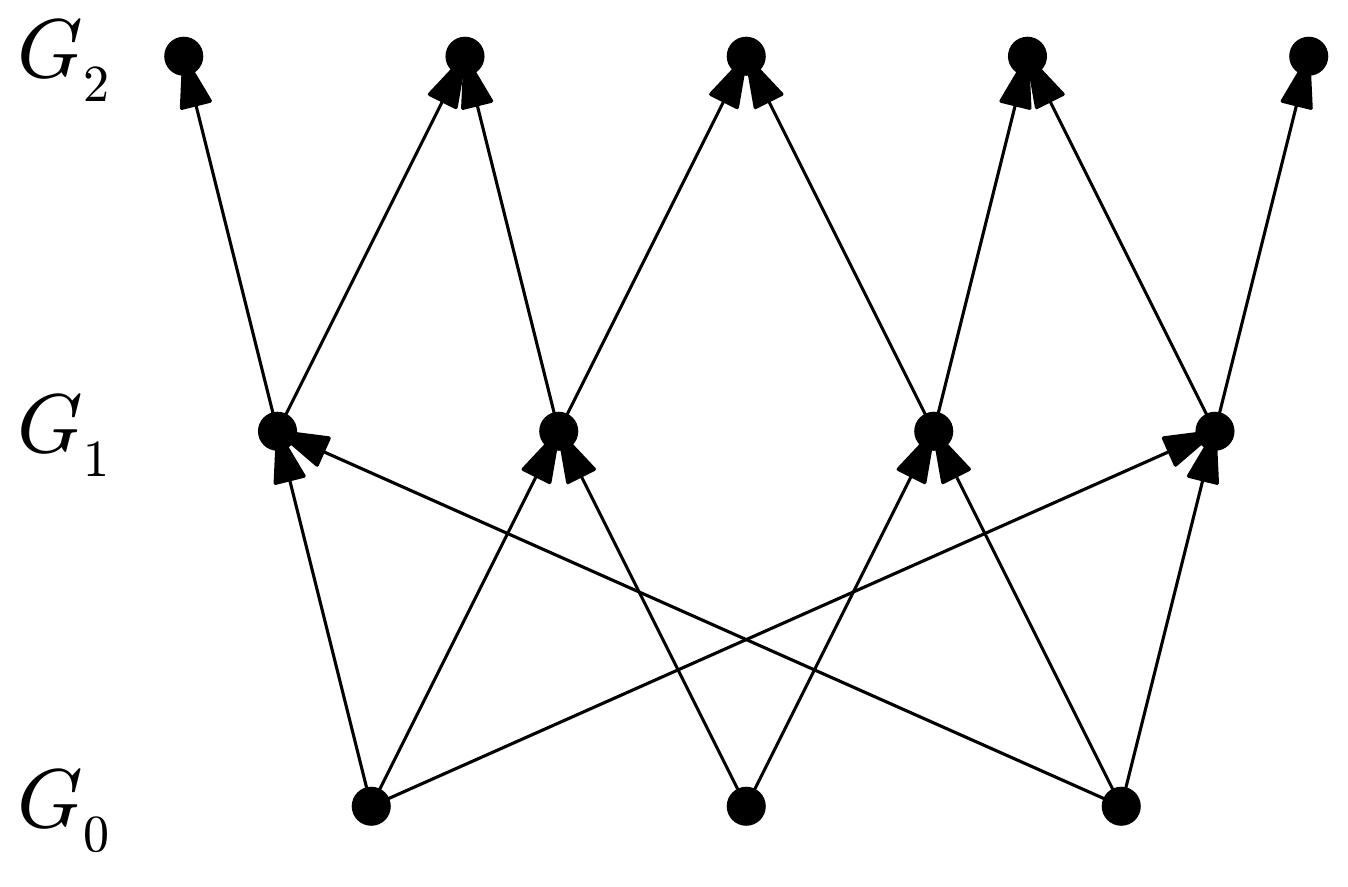,width=0.95\linewidth}
\caption{A growing causal set}
\label{fig:growcauset}
\end{center}
\end{minipage}
\end{figure}

Returning to the symmetry problem, and ignoring the general worry concerning Muller's fix based on weak discernibility above, the structuralist may therefore hope that she has been handed the necessary tools to deal with the symmetry problem. Upon inspection, however, one quickly notes that the irreflexivity of the one and only physical relation only affords a partial resolution of the problem. Suppose the causal set exhibits a high degree of {\em synchronic symmetry} in that every element in a given `horizontal' generation $G_i$ exemplifies exactly the same relational properties as any element of the same generation, as depicted in Figure \ref{fig:symcauset}, where points represent the elements of $C$ and the arrows between them the relation $\prec$ (rather than $\preceq$). This leads directly to a problem of {\em synchronic} plurality in that the PII suggests, once again, for causal sets relevantly like the one depicted in Figure \ref{fig:symcauset}, that each generation $G_i$ consists of just one element. It seems as if on a purely structuralist reading, and using PII, for all the generations $G_i$, all the points in it ought to be identified. This was essentially the symmetry problem as encountered for highly symmetric spacetimes in general relativity. Note that there is no analogous problem of {\em diachronic} plurality, since $\prec$ is asymmetric. It is of course no surprise that within a generation, the irreflexive relation does not dissolve the difficulty, as it cannot obtain between spacelike separated elements. 

It is worth noting that in slightly less symmetric cases, such as the one of an `expanding' universe as in Figure \ref{fig:growcauset}, not all points within a generation ought to be identified---even though there are subsets of points in each generation that share a relational profile. In generation $G_1$ in Figure \ref{fig:growcauset}, for instance, the leftmost and the rightmost points share all genuine physical properties and should thus be identified, in accordance with PII. The same applies to the two central points. Instead of the four points represented in $G_1$ in Figure \ref{fig:growcauset}, therefore, PII only identifies two events in the corresponding generation of the physical structure. This may serve as an indication that the strict symmetry is rather fragile in the sense that the vast majority of admissible causal sets do not exhibit symmetries challenging a structuralist interpretation.

Concerning potential problems of {\em diachronic} plurality, it was noted in \S\ref{ssec:kin} that the antisymmetry assumed in Axiom \ref{ax:partial} prohibits causal loops. But even if antisymmetry is dropped, a structuralist interpretation of causal set theory deprives the theory of the resources to deal with causal loops, for reason essentially the same as for the symmetric causal sets. 

To see why, let us suppose that antisymmetry is omitted from Axiom \ref{ax:partial}.\footnote{I wish to thank Fay Dowker for substantive discussions on this point.} We are then left with what is called a {\em transitive digraph}. If a causal set is merely a transitive digraph (rather than a partially ordered set), then there will be, in general, cycles in the sense that $\exists a, b, c \in C, a \prec b \prec c \prec a$. As a result of transitivity, any points $a$ and $b$ on the cycle will be relationally indistinguishable: $\forall x, y \in C,$ if $x\prec a$, then $x\prec b$ and vice versa, and if $a\prec y$, then $b\prec y$.\footnote{This is so because causal sets are `transitively closed' (cf.\ footnote \ref{fn:trans}).} Since points on loops can thus not differ in their relational profiles and the structuralist assumes that the elementary events of causal sets are intrinsically featureless, PII entails that all points on a loops are identical. But of course if all points on a loop are one and the same, then there is no loop, but just a lonely point. This means, quite generally, that transitive-digraph causal set theory cannot distinguish between causal loops and points in a causal set. Thus, either causal loops are ruled out {\em ab initio} (as they typically are, by assuming antisymmetry), or they immediately collapse to points, at least as long as we insist that elementary events have no intrinsic identity.

Since there are spacetimes with causal loops in general relativity and the points on these curves can be distinguished metrically (for instance by their distance to a fiducial point), no causal set could ever give rise to a spacetime-cum-causal loops, regardless of whether this causal set is conceived of as a partially ordered set or a transitive digraph. That a causal approach to discrete spacetime such as causal set theory cannot deal with closed timelike curves is not surprising, since these approaches are built upon Malament's (1977) insight that the metric is fixed, up to a conformal factor, by the causal structure---but only for spacetimes which satisfy a slightly stronger condition than that they do not contain causal loops.

Even though the symmetry problem thus resurfaces in causal set theory, the fragility mentioned three paragraphs ago indicates that the problem is not as grave as it is in general relativity. There, the symmetric spacetimes \`a la FLRW are arguably of measure zero in the space of all admissible spacetime models---even though the space of solutions to the Einstein field equations remains unknown. To take this as ground for dismissing the difficulty, however, was particularly unpalatable in general relativity since the FLRW models are of such great theoretical importance (W\"uthrich 2009). 

Returning to causal set theory, just how generic are these highly symmetric causal sets among all the kinematically possible ones? Even though this issue is hard to tackle analytically, one naturally conjectures that the fraction of highly symmetric causal sets quickly becomes very small as the cardinality of $C$ increases. Moreover, there is no reason to think that these highly symmetric possibilities are of particular physical relevance, as generic or non-generic these may be. In fact, there are grounds for thinking that they are not physically relevant: even FLRW and similarly symmetric spacetimes generically emerge from causal sets not exhibiting these perfect symmetries. That this is so is suggested by the fact that a faithful embedding of a causal set into a manifold can straightforwardly be obtained by randomly sprinkling points (at an appropriate density) into an FLRW spacetime and asserting causal relations between any pair of them according to the light cone structure. This will generate, {\em ex constructione}, a causal set that is faithfully embeddable into the FLRW spacetime. The random sprinkling, importantly, is required to guarantee the local Lorentz symmetry, as a sprinkling onto a regular lattice would not leave the average number of points in a volume invariant under Lorentz transformations.\footnote{In fact, the causal set depicted in Figure \ref{fig:symcauset} is not even manifold-like (Henson 2009, \S1.3).}

The fact that causal sets thus necessitate a certain irregularity gives the structuralist reassurance that the symmetry problem vanishes in causal set theory. This, in turn, indicates that the problem may disappear altogether in a full theory of quantum gravity.

\section{Zooming out: implications for structural realism}\label{sec:impl}

By way of conclusion, I wish to step back and take (admittedly preliminary) stock of the situation for the structuralist with respect to fundamental theories of gravity, as well as more generally in philosophy of science. In order to do that, let me distinguish two different structures that might be relevant here. First, a particular model of a theory is a mathematical structure that is claimed, by the theory, to represent a physically possible world according to that theory. Second, the theory itself, i.e.\ the set of models, has a structure. This structure, too, is a mathematical structure encoded in the mathematical formulation of the theory. The mathematical structure of the theory and that of its models may of course be related, but they need not be. Applied to causal set theory, the structure of a model of the theory is the structure of a particular causal set as explicated in the previous section.\footnote{In fact, we may distinguish between kinematic and dynamic models; the former being just causal sets, while the latter would be sequences of causal sets.} The structure of causal set theory, on the other hand, is captured by the structure of the set of kinematically admissible causal sets, i.e.\ of the causal sets complying with Axioms \ref{ax:partial} and \ref{ax:discrete}, or perhaps by the partially ordered structure $\mathcal{P}$ structure of dynamically possible causal sets, i.e.\ of sequences of causal sets conforming to the sequential growth dynamics as explicated in \S \ref{ssec:dyn}.\footnote{One may reasonably think that any theory purporting to recover general relativity in some limit could only succeed in doing so if appropriate dynamical aspects of the latter are regained as well. The role of dynamics in understanding just how general-relativistic spacetimes emerge from causal sets defies, so far, satisfactory explication. This is a restatement of the problem mentioned, and deferred, in footnote\ref{fn:dyn}. I thank an anonymous referee for pressing me on this point.}

Naively, a theory that purports to recover general relativity in a classical limit should do more than catalog kinematically admissible states.  One would think that dynamics is essential in this endeavor. 

Instead of the more ambitious goal of offering a wholesale solution to negotiate the competing demands of realist and antirealist arguments to which Worrall (1989) aspired, we may more modestly seek to formulate retail interpretations of particular physical theories, one by one. What has been offered above was intended in the spirit of the latter. This more modest goal is both more promising and more appropriate.\footnote{I am not alone in this view; cf.\ Esfeld and Lam (2008, 2011).} More promising because in order to make headway on the first ambition, we are faced with a number of apparently insurmountable obstacles, while the latter goal can be, and in some cases, has been, achieved. More appropriate since there is no guarantee for any universal attitude towards scientific theories in general that it will succeed in that either it correctly licenses the theories' epistemic pretensions or at least furthers our understanding of science. 

Putting aside modesty for a moment, what would be required for a structural realist to successfully steer clear of the antirealist threat arising from past false theories? About which of the structures identified in the first paragraph of this section ought we be realist? Answering this second question first, it seems clear that the realism should be directed at the structure of particular models rather than of the theory and hence of the entire set of models. Unless we assume that the sum total of existence consists of a set of physically possible worlds the structure of which our best scientific theory must reproduce, unless, that is, we are some sort of modal realists and our modal realism dictates our philosophy of science, I see little motivation to be realist about the entire abstract structure of the mathematical formulation of the theory at stake. This is a rather deep point, a point that has recently been come under pressure from various quarters, and a point that most certainly deserves further consideration than it can be given here. 

But if I am right in rejecting a need for realism concerning the entire structure of a theory, then even a wholesale structural realism should focus its realism on particular model, or perhaps models, that are thought to represent aspects of the world. Since the realism is structural, the claim must be that the model(s) correctly represent(s) the structure of the physical world. In the now standard terminology of Shapiro (1997), the structural realist must thus be an {\em in re} structuralist, i.e.\ must maintain that the abstract structure of the model is instantiated in the physical world. In other words, the structural realist posits an isomorphism between the model(s) and the aspects of the world represented. We should of course never expect that any model of any scientific theory is fully isomorphic to the world. Rather, the claim should be that there exist substructures in the world and that the model is isomorphic to that substructure or that there are several models of the theory, all of which are isomorphic to different substructures of the world.\footnote{A {\em substructure} $\mathcal{T}$ of a structure $\mathcal{S} = \langle S, R\rangle$ is an ordered pair $\langle T, R|_T\rangle$ consisting of a domain $T\subset S$ and the set $R$ of relations restricted to $T$.} Of course, the challenge is to correctly identify the substructure in the world with which the model is supposed to be isomorphic, all the while avoiding the risk of trivializing the isomorphic achievement by insisting on some criterion of explanatory strength. I leave the details of this tricky, but crucial, step to some other occasion.

Even in the absence of a principled account of how to identify the pertinent substructures in the world, for the wholesale structural realist to meet the antirealist challenge, there must be isomorphisms between substructures of the models of succeeding theories in the relevant sense in order to underwrite the necessary structural continuity across scientific revolutions. While once again I see, to put it mildly, little hope to fulfill the ambitions of establishing the relevant isomorphisms quite in general, abstracted away from particular scientific theories, progress may be achieved by studying particular cases of transitions from one theory to its successor. Above, we identified good reasons not to take causal set theory as offering a true fundamental theory of quantum gravity and instead to consider it a toy theory which may reveal some aspects of such a theory. For the sake of illustrating the point, pretend it {\em did} offer such a full theory and supplanted general relativity as our best theory of gravity. The task, then, is to show this structural continuity in the imagined transition from general relativity to causal set theory. 

Again, it is clear that models in the two theories cannot be fully isomorphic. First, general-relativistic models are much richer structures with their domains being the points of four-dimensional differentiable, and therefore continuous, manifolds. This means that there is a mathematically precise sense in which there are many more elementary events than in a causal set. Additionally, these `extra' events also exemplify physical relations, thus dashing any hope of full isomorphism. Moreover, general-relativistic structures contain more relations, including those arising from the Einstein field equations, than do those of causal set theory. If there is structural continuity between the two, then it must manifest itself in the form of partial isomorphisms between their models, i.e.\ of isomorphisms between causal sets and substructures of the general-relativistic spacetimes. This brings us back to the problem of how causal sets give rise to spacetimes discussed in \S \ref{ssec:emerge}.

There, we have noted that most causal sets are not manifold-like and that it remains unclear how there could be dynamical or otherwise physical mechanisms that choose those causal sets which do indeed give rise to classical spacetimes. If we believe that general relativity is a successful theory insofar as it contains models which faithfully describe most aspects of spatio-temporal structure as we find it in the actual world---and so we should---, then any quantum theory of gravity will have to let those models emerge in a principled manner. I will not offer a solution to this problem today, but hasten to insist that solving it will mean to address the wanted structural continuity between the models of the two theories.

The example of the relationship between causal sets and relativistic spacetimes thus illustrates how particular cases of related theories, such as those related by succession, may be much more fruitful than any aggrandizing general argument as to how successive theories must in general relate to one another: by identifying which causal sets may be structurally related to relativistic spacetimes, we may hope to elucidate the problem of emergence addressed in \S\ref{ssec:emerge}. 

In conclusion, let me emphasize again that structuralism may proffer a natural interpretation of particular physical theories such as causal set theory, and perhaps even shed light on particular instances of transitions between succeeding theories. Little can be found, however, in causal set theory and how it relates to relativistic spacetime theories that would rekindle hope for using structuralism as a fully general template in the philosophy of science.


\begin{thebibliography}{99}

\bibitem{bai06} Bain, J.\ (2006). Spacetime structuralism. In D.~Dieks (ed.), {\em The Ontology of Spacetime} (pp.\ 37-66). Amsterdam: Elsevier.

\bibitem{bai11} Bain, J.\ (forthcoming). Category-theoretic structure and radical ontic structural realism. {\em Synthese}.

\bibitem{bainor} Bain, J.\ and Norton, J.D. (2001). What should philosophers of science learn from the history of the electron? In J.Z.~Buchwald and A.~Warwick (eds.), {\em Histories of the Electron: The Birth of Microphysics} (pp.\ 451-465). Cambridge, MA: MIT Press.

\bibitem{bomeal97} Bombelli, L., Lee, J., Meyer, D.\ and Sorkin, R.D.\ (1987). Space-time as a causal set. {\em Physics Review Letters} {\bf 59}: 521-524.

\bibitem{brieal02} Brightwell, G., Dowker, H.F., Garc\'ia, R.S., Henson, J.\ and Sorkin, R.D.\ (2002). General covariance and the `problem of time' in a discrete cosmology. In K.~Bowden (ed.), {\em Correlations: Proceedings of the ANPA 23 Conference, August 16-21, 2001, Cambridge, England} (pp.\ 1-17). London: Alternative Natural Philosophy Association.

\bibitem{but07} Butterfield, J.\ (2007). Stochastic Einstein locality revisited. {\em British Journal for the Philosophy of Science} {\bf 58}: 805-867.

\bibitem{dow05} Dowker, F.\ (2005). Causal sets and the deep structure of spacetime. In A.~Ashtekar (ed.), {\em 100 Years of Relativity: Space-Time Structure: Einstein and Beyond} (pp.\ 445-464). Singapore: World Scientific.

\bibitem{doweal04} Dowker, F., Henson, J.\ and Sorkin, R.D.\ (2004). Quantum gravity phenomenology, Lorentz invariance and discreteness. {\em Modern Physics Letters} A{\bf 19}: 1829-1840.

\bibitem{ear08} Earman, J.\ (2008). Reassessing the prospects for a growing block model of the universe. {\em International Studies in the Philosophy of Science} {\bf 22}: 135-164.

\bibitem{esflam08} Esfeld, M.\ and Lam, V.\ (2008). Moderate structural realism about space-time. {\em Synthese} {\bf 160}: 27-46.

\bibitem{esflam11} Esfeld, M.\ and Lam, V.\ (2011). The structural metaphysics of quantum theory and general relativity. Manuscript.

\bibitem{fre10} French, S.\ (2010). The interdependence of structure, objects, and dependence. {\em Synthese} {\bf 175}: 89-109 (supplement). 

\bibitem{geo05} Georgiou, N.\ (2005). The random binary growth model. {\em Random Structures and Algorithms} {\bf 27}: 520-552.

\bibitem{haweal76} Hawking, S.W., King, A.R.\ and McCarthy, P.J.\ (1976). A new topology for curved space-time which incorporates the causal, differential, and conformal structures. {\em Journal of Mathematical Physics} {\bf 17}: 174-181.

\bibitem{hen09} Henson, J.\ (2009). The causal set approach to quantum gravity. In D.~Oriti (ed.), {\em Approaches to Quantum Gravity: Toward a New Understanding of Space, Time and Matter} (pp.\ 393-413). Cambridge: Cambridge University Press.

\bibitem{hof79} Hooft, G.\ 't (1979). Quantum gravity: A fundamental problem and some radical ideas. In M.~L\'evy and S.~Deser (eds.), {\em Recent Developments in Gravitation: Carg\`ese 1978} (pp.\ 323-345). New York and London: Plenum Press.

\bibitem{jon91} Jones, R.\ (1991). Realism about what? {\em Philosophy of Science} {\bf 58}: 185-202.

\bibitem{lad98} Ladyman, J.\ (1998). What is structural realism? {\em Studies in the History and Philosophy of Modern Physics} {\bf 29}: 409-424.

\bibitem{lad09} Ladyman, J.\ (2009). Structural realism. In E.N.~Zalta (ed.), {\em Stanford Encyclopedia of Philosophy}.

\bibitem{ladeal} Ladyman, J.\ and Ross, D., with Spurrett, D.\ and Collier, J.\ (2007). {\em Every Thing Must Go: Metaphysics Naturalized}. Oxford: Oxford University Press.

\bibitem{lamwut} Lam, V., and W\"uthrich, C. (2012). No categorical support for radical ontic structural realism. Manuscript.

\bibitem{leilad} Leitgeb, H.\ and Ladyman, J.\ (2008). Criteria of identity and structuralist ontology. {\em Philosophia Mathematica} {\bf 16}: 388-396.

\bibitem{lan07} Landry, E.\ (2007). Shared structure need not be shared set-structure. {\em Synthese} {\bf 158}: 1-17.

\bibitem{mal} Malament, D.B.\ (1977). The class of continuous timelike curves determines the topology of spacetime. {\em Journal of Mathematical Physics} {\bf 18}: 1399-1404.

\bibitem{mul10} Muller, F.A.\ (2010). The characterization of structure: definition versus axiomatisation. In F.~Stadler et al.\ (eds.), {\em The Present Situation in the Philosophy of Science} (pp.\ 399-416). Berlin: Springer.

\bibitem{mul11} Muller, F.A.\ (2011). How to defeat W\"uthrich's abysmal embarrassment argument against space-time structuralism. {\em Philosophy of Science} {\bf 78}: forthcoming.

\bibitem{myr78} Myrheim, J.\ (1978). Statistical geometry. CERN preprint TH-2538, available at http://doc.cern.ch/archive/electronic/kek-scan/197808143.pdf.

\bibitem{poo06} Pooley, O.\ (2006). Points, particles, and structural realism. In Rickles et al.\ (pp.\ 83-120).

\bibitem{rei01} Reid, D.D.\ (2001). Discrete quantum gravity and causal sets. {\em Canadian Journal of Physics} {\bf 79}: 1-16.

\bibitem{riceal06} Rickles, D., French, S.\ and Saatsi, J.\ (eds.) (2006). {\em The Structural Foundations of Quantum Gravity}. Oxford: Oxford University Press.

\bibitem{ridsor99} Rideout, D.P.\ and Sorkin, R.D.\ (1999). Classical sequential growth for causal sets. {\em Physical Review} D{\bf 61}: 024002.

\bibitem{ridwal09} Rideout, D.\ and Wallden, P.\ (2009). Emergence of spatial structure from causal sets. {\em Journal of Physics: Conference Series} {\bf 174}: 012017.

\bibitem{rob10} Roberts, B.\ (2011). Group structural realism. {\em British Journal for the Philosophy of Science} {\bf 62}: 47-69.

\bibitem{sau03} Saunders, S.\ (2003). Structural realism, again. {\em Synthese} {\bf 136}: 127-133.

\bibitem{sha97} Shapiro, S.\ (1997), {\em Philosophy of Mathematics: Structure and Ontology}. New York: Oxford University Press. 

\bibitem{smewut} Smeenk, C.\ and W\"uthrich, C. (2011). Time travel and time machines. In C.~Callender (ed.), {\em The Oxford Handbook in the Philosophy of Time} (pp.\ 577-630). Oxford: Oxford University Press.

\bibitem{smo06} Smolin, L. (2006). The case for background independence. In Rickles et al.\ (pp.\ 196-239).

\bibitem{sta06} Stachel, J.\ (2006). Structure, individuality, and quantum gravity. In Rickles et al.\ (pp.\ 53-82).

\bibitem{varrid06} Varadarajan, M.\ and Rideout, D.P.\ (2006). A general solution for classical sequential growth dynamics of causal sets. {\em Physics Review} D{\bf 73}: 104021.

\bibitem{wor87} Worrall, J.\ (1989). Structural realism: the best of both worlds? {\em Dialectica} {\bf 43}: 99-124.

\bibitem{worzah} Worrall, J.\ and Zahar, E.\ (2001). Ramsification and structural realism. Appendix in E.~Zahar, {\em Poincar\'e's Philosophy: From Conventionalism to Phenomenology} (pp.\ 236-251). Chicago and La Salle, IL: Open Court.

\bibitem{wut09} W\"uthrich, C.\ (2009). Challenging the spacetime structuralist. {\em Philosophy of Science} {\bf 76}: 1039-1051.

\end{thebibliography}
\end{document}